\documentclass{article}
\usepackage{arxiv}
\usepackage[utf8]{inputenc} % allow utf-8 input
\usepackage[T1]{fontenc}    % use 8-bit T1 fonts
\usepackage{url}            % simple URL typesetting
\usepackage{booktabs}       % professional-quality tables
\usepackage{amsfonts}       % blackboard math symbols
\usepackage{nicefrac}       % compact symbols for 1/2, etc.
\usepackage{microtype}      % microtypography
\usepackage{lipsum}
\usepackage{graphicx}
\graphicspath{ {./images/} }
\usepackage{moreverb,url}
\usepackage[colorlinks,bookmarksopen,bookmarksnumbered,citecolor=red,urlcolor=red]{hyperref}
\usepackage{natbib}

\newcommand\BibTeX{{\rmfamily B\kern-.05em \textsc{i\kern-.025em b}\kern-.08em
T\kern-.1667em\lower.7ex\hbox{E}\kern-.125emX}}

\title{Integrating smart surveys with traditional surveys}

\author{
 Danielle McCool \\
  department of Methodology and Statistics\\
  Utrecht University\\
  \texttt{d.m.mccool@uu.nl} \\
   \And
 Peter Lugtig \\
  department of Methodology and Statistics\\
  Utrecht University\\
  \texttt{p.lugtig@@uu.nl} \\
  \And
 Bella Struminskaya\\
  department of Methodology and Statistics\\
  Utrecht University\\
  \texttt{b.struminskaya@uu.nl} \\
}
\begin{document}
\maketitle

\begin{abstract}
Smart surveys are surveys that   make use of sensors and machine intelligence to reduce respondent burden and increase data quality. Smart surveys have been tests as a way to improve diary surveys in official statistics, where data are collected on topics such as travel, time use and household expenditures. There are often inherent differences both in measurement and representation between smart surveys and traditional diaries, which makes it difficult to integrate both data sources in producing statistics over time, or within a mixed- or multi-source context. 
This paper distinguishes two different approaches to integration: the mixed-mode approach, which prioritizes outcome alignment and minimizes measurement differences for straightforward data merging, and the multisource approach, which maintains inherent mode differences and integrates data at the modeling stage, allowing exploitation of the strengths of each source. Using travel surveys as an illustrative example, we explore the benefits and drawbacks of each approach, and propose a decision framework to guide researchers in selecting the appropriate integration strategy.

\end{abstract}


\section{Introduction}
Smart surveys offer a promising advancement in data collection by leveraging features available in modern smartphones \textemdash such as sensors and machine intelligence \textemdash to help survey respondents report on complex or intensive behaviors. In the context of official statistics, complex behaviors such as mobility habits, time use, or expenditures are routinely collected, traditionally using diaries \citep{Adler2002-yx,astin2021measuring,hozer2016examining}. Respondents don’t enjoy (and are therefore often poor at) tasks that require continuous reporting, or that rely on recall of detailed behaviors \citep{Husebo2018-cw, Johnston2014-nq, Mayer2019-rf}. Conveniently, computers are generally well suited for these tasks, a fact which has not gone unnoticed by survey researchers. The last decade has been marked by an increase in smart surveys, which seek to augment existing methodology by using the tools readily available within smartphones \citep{Couper2018-ct, Link2014-zj, Struminskaya2020-ij}. Automating parts of the data capture can potentially reduce respondent burden and increase accuracy, leading to improved data quality. 

The adoption of smart surveys introduces the challenge of how to integrate smart survey data with data from more traditional survey methods. Measurement differences are likely to arise between the two methods due to inherent differences in data collection methods. Paper or web-based diaries have traditionally relied on self-reported measures. Respondents keep track of all their trips, activities and expenditures, and write these down over a period of a few days or weeks.  Smart surveys try to assist the respondent by making use of sensors and machine intelligence. For example, GPS sensors can be used to track a respondents location over time \citep{McCool2021-ae}, accelerometers can be used to measure activities \citep{hohne2020motion}, and the camera can be used to make pictures of (receipts of) expenditures \citep{schouten2025can}.

Although smart surveys can achieve relatively good response rates \citep{Lugtig2022-ku}, there are reasons to believe that smart surveys suffer from problems with representativity. One self-evident example is that a person without a smartphone can’t participate in a smart survey. While this specific group may be shrinking, with a report from the Eurobarometer indicating that in 2021 96\% of Europeans report had access to a mobile phone \citep{noauthor_2021-mj}), there remains ample evidence demonstrating that both smartphone ownership and smartphone usage are unevenly distributed within the population \citep{Keusch2023-tg, Klingwort2020-al}. Even where there is access to a suitable device, privacy concerns may reduce participation in some groups \citep{Roberts2022-hb}, and age and educational background may also be associated with reduced response rates in smart surveys \citep{Lugtig2022-ku}. Over time, these issues may become less relevant, as they have for other survey modes, but even researchers who opt out of a mixed-mode of administration for their survey for reasons of representation must address the issue that a switch towards smart surveys may cause a break in the time-series of statistics. Even when smart surveys would be used as a standalone method in data collection, there is the issue of how to analyse statistics longitudinally. Method changes may lead to breaks in the time-series of statistics that are undesired.

Researchers have grappled with these questions for decades in the context of mixed-mode survey design, in which the same content was administered via differing modes (for example telephone versus web) to access different groups of people to increase response rates by capturing groups of people who may be more likely to respond via a particular mode. To accomplish this, most of the methods explicitly aim to eliminate differences in measurement between the modes as much as possible \citep{Burton2020-tr, Klausch2013-xr}. Question wording is revised as necessary, layouts adapted, and instructions reworked to align response categories between modes as closely as possible. While different survey modes can still lead to some measurement differences (e.g. for  sensitive behaviors \citep{kreuter2008social}), measurement differences in traditional mixed-mode surveys are often small, allowing integration by simply combining data from multiple modes into a single dataset \citep{Antoun2019-nw, Couper2017-ea}.

\begin{figure}
     \centering
     \includegraphics[width=1\linewidth]{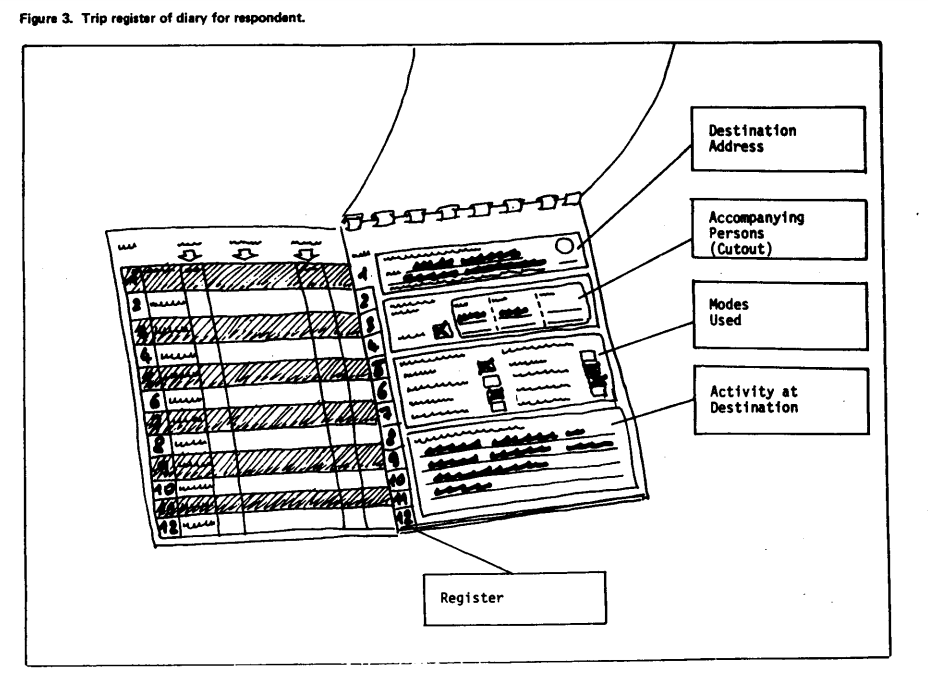}
     \caption{An example early mobility diary requiring a respondent to fill in details about their travel behavior. Reproduced from \cite{Ampt1985-gf}.}
     \label{fig:paperdiary}
 \end{figure}

Some of the largest gains in terms of better measurement for smart surveys come from household diary studies. As an example, consider the mobility diary in Figure~\ref{fig:paperdiary} in which respondents are asked to record their travel behavior including mode of transportation, precise start-, and stop times for each trip, and addresses for each place visited. Traditionally, travel surveys have been conducted in a paper diary format, with each row of the diary containing details of each successive trip and each page spanning a new day. The duration of diary entries have ranged from a full 24 period to several weeks, depending on the study \citep{Axhausen1995-dn}. Trip underreporting, especially with respect to short trips, has been an issue with these diaries from the beginning \citep{Ampt1985-gf, Ashley2009-zw}. 

 Researchers have been looking into ways to improve travel-diary studies with smart features like stand-alone GPS loggers and form-filling features in web-based versions since the 1990s \citep{Adler2002-yx, Bricka2009-fr, Sarasua1996-vr}. Once location-tracking and internet connectivity became standard in smartphones, deploying smart surveys onto personal mobile devices was a logical follow-up \citep{Berger2015-ec, Cottrill2013-ym, Greaves2015-on, Nitsche2014-na}. App-recorded data are generally more accurate than self-reported data; times are more precise, short trips are more likely to be recorded, and distance estimation is better \citep{McCool2021-ae}. Smart survey data are moreever more granular than the traditional diary data, providing an inroad for answering new research questions. For example, in a smart survey it is possible for most recorded trips  to determine the exact route that a respondent took, as well as the speed on different sections. 

This improvement in measurement in smart surveys comes at a cost: the data are no longer so similar in nature that they can be naively integrated with non-smart data such as paper diaries in one dataset. A logical question might be whether or not this integration with non-smart modes is even necessary -- there would seem to be nothing lost in the decision to replace the old mode with a new mode that is categorically better. In such a case, the role of integration might be temporary, to bridge gaps over time and allow for comparisons in long-running research. In other cases, however, data integration is likely to be a perennial concern for smart surveys, which are rarely categorically better with respect either to measurement or representation. Data integration is an unavoidable concern.

We propose in this paper that there are two fundamentally different approaches to the question of data integration of smart- and traditional data: the mixed-mode approach and the multisource approach. The mixed-mode approach prioritizes minimization of mode measurement error, adjusting one mode to accommodate another as a form of pre-processing. The multisource approach prioritizes minimization of overall measurement error, integrating the data at the modeling stage. Section~\ref{sec:twotypes} of this paper describes both methods, including potential benefits and drawbacks. Sections~\ref{sec:mixedmode} and \ref{sec:multisource} detail the mixed-mode and multi-source approaches respectively, using the mobility app as an illustrative example. Finally, Section~\ref{sec:discussion} provides recommendations for researchers looking to bridge the gap between smart and non-smart surveys.

\section{Two types of integration} \label{sec:twotypes}

The overall goal of a smart survey is to improve the quality of data, most often by improving measurement. It might do this by reducing the burden for respondents, leading to less satisficing or non-response, or by measuring things that aren’t particularly easy for respondents to gauge, such as the distance they walked on a particular trip. 
 
For example, consider the development of an online mobility diary meant to function as the primary mode in ongoing travel research. Statistics Netherlands has measured the mobility of the Dutch population since 1978 using diary studies. The diaries have over time evolved from face-to-face interviews, to various versions of a mixed-mode survey including telephone, mail and web questionnaires,  finally resulting in a web-only diary survey \citep{Boonstra2019-uz}. The individual travel diary was originally designed for a pencil and paper diary (PAPI), and the later web version was developed such that it aligned as closely as possible with the layout and question formatting of the original paper diary. 

\begin{figure*}
    \centering
    \includegraphics[width=1\linewidth]{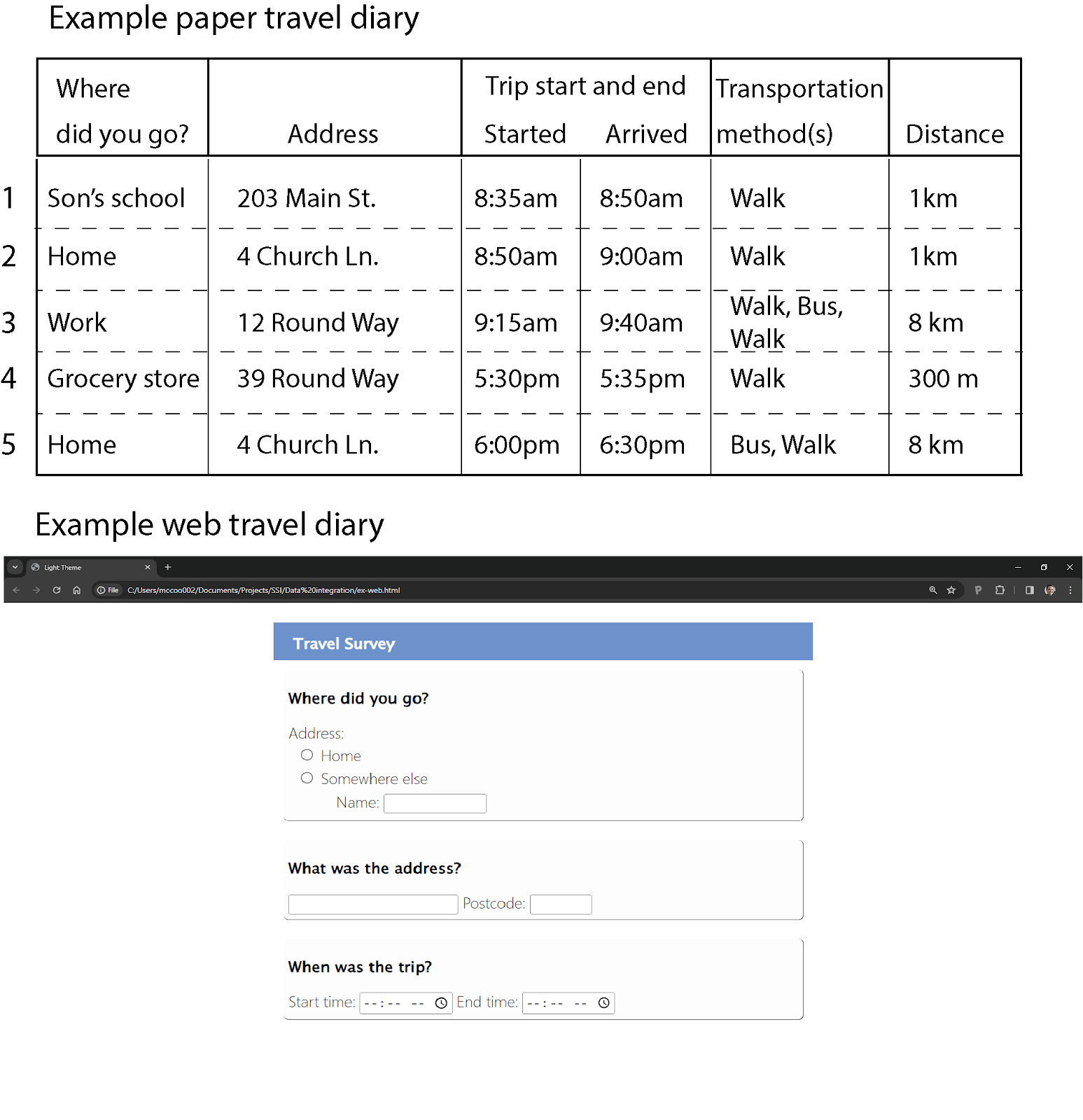}
    \caption{An example paper diary and web travel diary developed to minimize differences in incoming data.}
    \label{fig:paperweb}
\end{figure*}

Figure~\ref{fig:paperweb} illustrates a mixed-mode travel questionnaire, where some respondents will complete the questionnaire on the web, and others on paper. The alignment of the content is constrained such that respondents are asked the same questions in as near an identical way as possible over the two different modes. The start and end times, transport method and locations of trips are asked of the respondent across both modes. The data resulting from both instruments, once in tabular form, will look very much the same. Each column in the paper diary or question in the web survey contributes descriptive data about a period of staying in the same place, or a trip. Rows within an individual’s dataset delineate distinct trips. If the same day were reported perfectly in both modes, the data would in theory be identical.

\begin{figure}
    \centering
    \includegraphics[width=.5\linewidth]{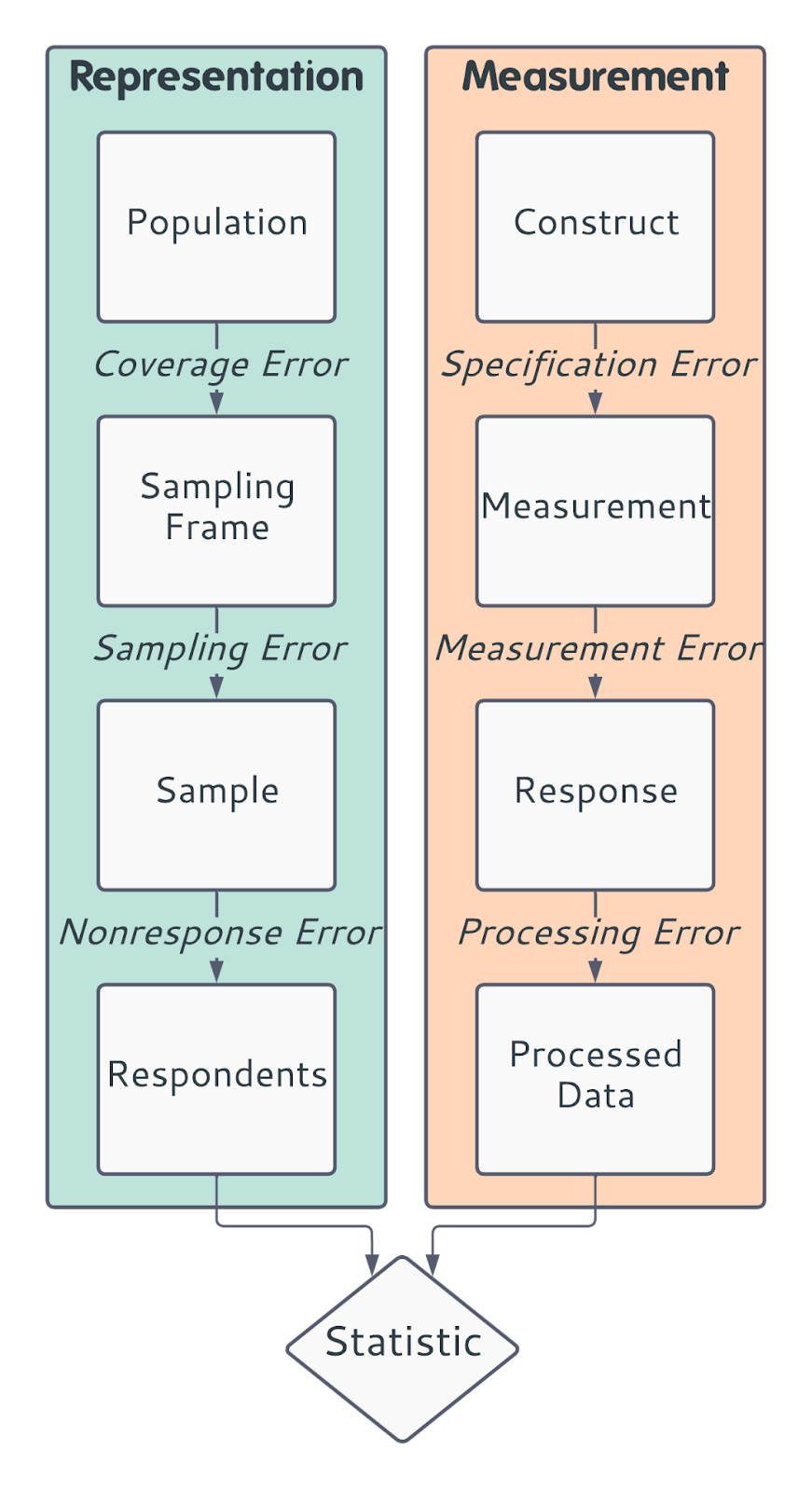}
    \caption{The TSE model, split into Representation and Measurement components. The mixed-mode paradigm for data integration is concerned with reducing differences within Measurement in order to address differences in Representation.}
    \label{fig:tse}
\end{figure}

 In practice, measurement error in the completion of trip details arises in both modes. Respondents may forget some trips, round the duration of trips, or give an imprecise location in similar ways on the web and on paper.  Figure~\ref{fig:tse} illustrates the different components of the Total Survey Error model as proposed by \cite{Groves2010-ik}, split into error components of representation and measurement. If we assume that two modes don’t differ on the measurement side -- that the specification error, measurement error, and processing error is similar in both modes -- then it’s possible to focus entirely on differences in representation \citep{Vannieuwenhuyze2013-dr}. We may comfortably attribute any differences in outcomes as being consequent to these representational differences. If web respondents report more trips than Paper-and-pencil respondents, we can attribute this difference to the fact that different types of respondents answer to each survey mode. This general concept forms the basis of the mixed-mode type of integration: when measurement differences are negligible, data from two different modes can be combined in one dataset, and in further analyses, the survey mode can be ignored. In the case when we have small differences between modes, a similar approach can be followed. Data from the different modes are combined in one dataset, but some calibration between modes is usually conducted to control for measurement differences\citep{Vannieuwenhuyze2014-tu}. After calibrations, data integration is completed.
 
The second way to integrate data is by following the framework of multi-source statistics. The underlying idea of multisource statistics, is to analyse data stemming from multi-sources separately, and only integrate the data at the very end \citep{Waal2020-ii}. We will argue in this paper that such an approach is preferable in situations where there are large differences in the measurement side of the TSE model shown in figure~\ref{fig:tse}.

\begin{figure*}
    \centering
    \includegraphics[width=1\linewidth]{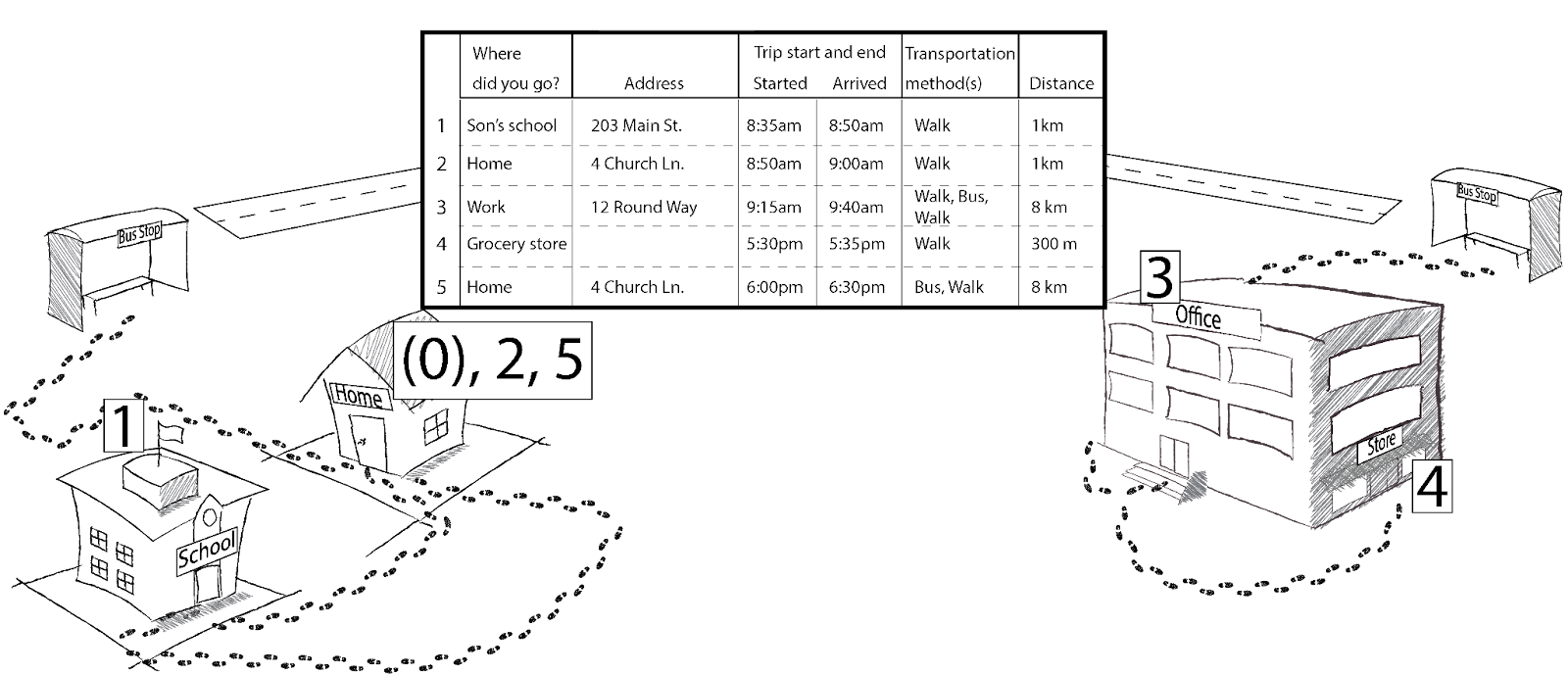}
    \caption{Paper diary representation of underlying travel behavior.}
    \label{fig:papreal}
\end{figure*}

Smart surveys will in many situations do measurements in a much more granular way, and  for that reason introduce intentional differences in the measurement side in Figure~\ref{fig:paperweb}For example,a smart survey may reduce the measurement error in estimating travel distance by using GPS tracks to measure the actual travel distance, rather than the inferred distance as is done in a traditional diary. A smart survey using GPS sensors will consist of long sequences of GPS points rather than the begin- and end-locations as in a travel diary survey.
Expanding the example from Figure~\ref{fig:paperweb}, we might imagine the true underlying travel behavior of this respondent to be as shown Figure~\ref{fig:papreal}. At some point in time, the respondent leaves his or her home walks, following a certain route to a school where he or she lingers briefly before returning home, and then initiating a commute to work. When these traces are captured in a paper diary, it will likely contain some inaccuracies concerning trip start and end times or travel distances. Other information, such as the route choice, or time allocated per mode of transportation, will not be present at all.

\begin{figure*}
        \centering
    \includegraphics[width=1\linewidth]{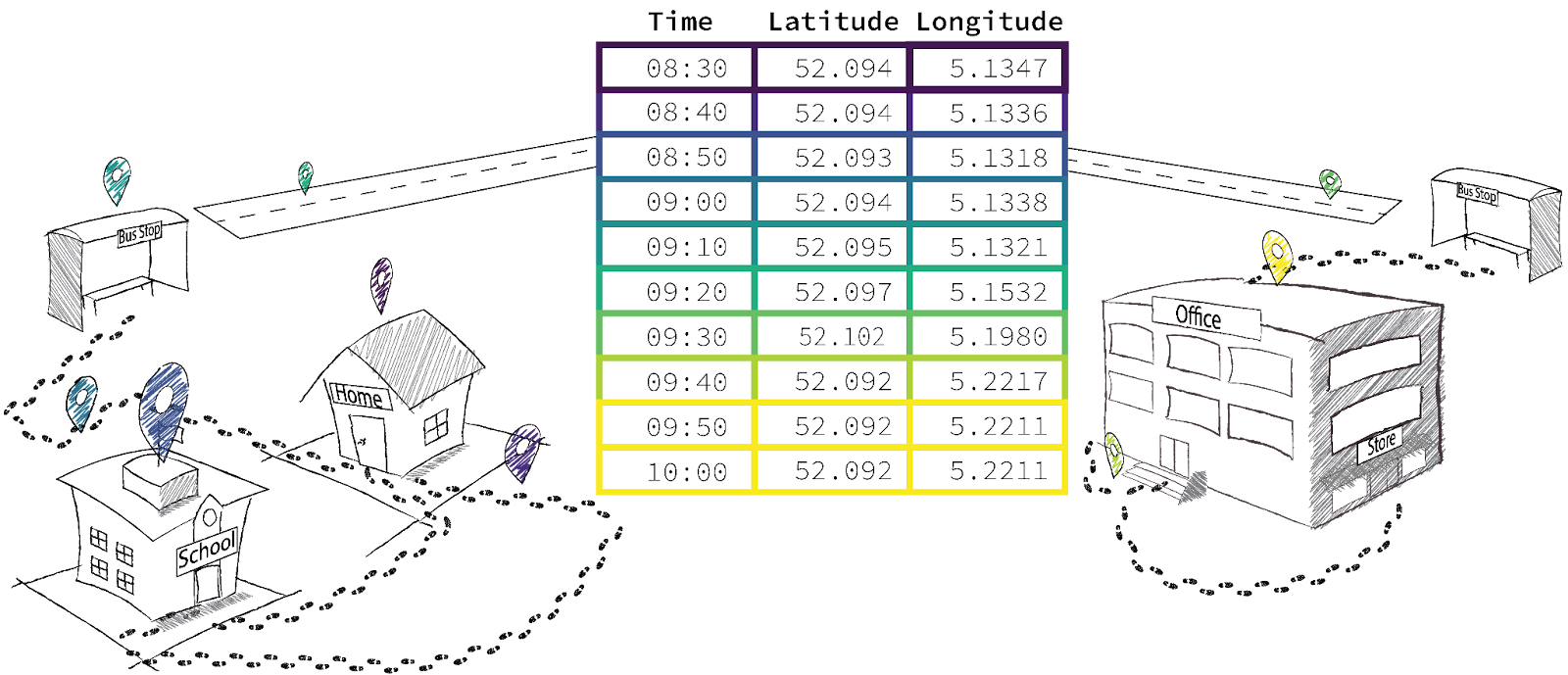}
    \caption{Sensor measurements of the same underlying travel behavior as part of a smart survey.}
    \label{fig:sensreal}
\end{figure*}

Figure~\ref{fig:sensreal} illustrates what location data are captured by the sensor component of a smart survey for the same trip. The path to integrating these raw data -- consisting of time-stamped latitudes and longitudes -- with data generated by a traditional survey is not straightforward. 

In this case, a researcher who is interested in integrating or comparing smart and non-smart surveys for this travel behavior is forced to make a decision on how best to address measurement differences betweeen the paper diary survey and GPS-based measures from the smart survey. There are two options: 1) constrain the smart survey data applying a set of algorithms to get the data into the same shape as the traditional diary data, or 2) leave the inherent different data structure of smart surveys intact, produce statistical estimates in the smart and traditional diary separately based on two different datasets with different data processing models, and only integrate the statistical estimates at the very end of the study.

The following section describes what we call the mixed-mode approach in which the raw data captured within a smart survey are processed before integration. We describe this method of integration using a hypothetical travel survey for illustration, and outline the benefits and drawbacks of this method. 

\section{Mixed-mode-like integration using travel studies as example}\label{sec:mixedmode}

The ultimate goal of mixed-mode-like integration is close alignment of the two data sources, leading to a single integrated dataset. To this end, a researcher will seek to reduce the measurement differences between a smart and non-smart survey. This may require ignoring structural aspects of the smart survey data that improve measurement, aggregating across more coarse intervals of time, or removing incomplete measurements.

\begin{figure*}
    \centering
    \includegraphics[width=1\linewidth]{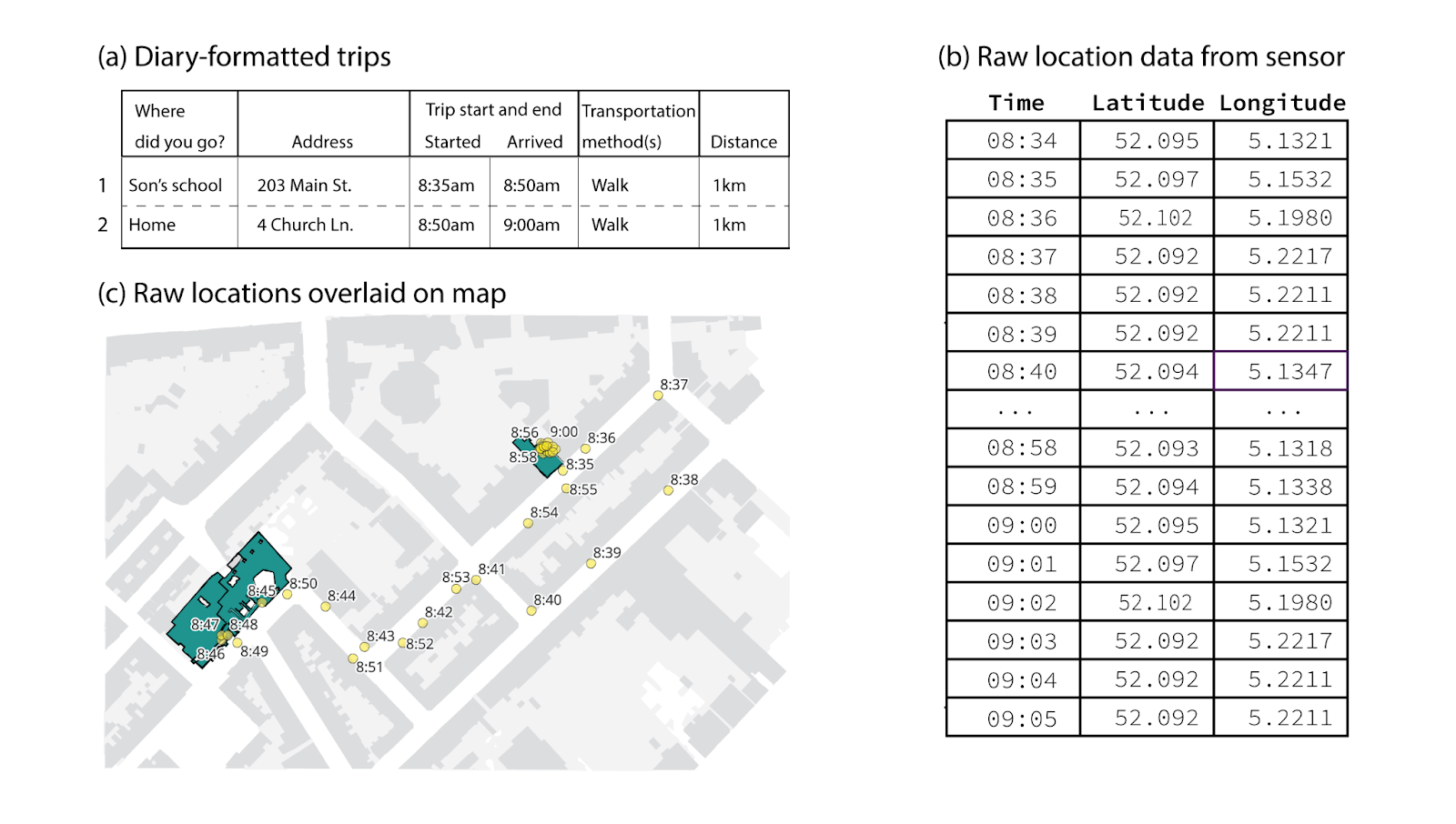}
    \caption{Two trips, represented in (a) diary format, (b) raw location data format, and (c) as time stamped locations overlaid on a map. To create diary-formatted data from the raw location data requires substantial processing. }
    \label{fig:diaryvsraw}
\end{figure*}

We can use the mobility example from Section 2 to illustrate the process. In Figure~\ref{fig:diaryvsraw}, we focus on the first two trips from the travel diary: a parent walking to and from their child’s school. When this data is represented in diary format, each trip, represented as a row, is fully described by six fields: the label/context of the location, the address, the start time, the stop time, the transportation method or methods, and the distance traveled. If we were to record a person’s location coordinates over the same period of time in a smart survey, a trip could comprise any number of rows, depending on how often the location is sampled on the device, and the actual duration of the trip. In the simplest case, each row in the data is represented by only three columns: latitude, longitude, and a timestamp. To bring this smart data data into the same format as the traditional diary means that the six fields from the diary format must be derived from the series of location coordinates and timestamps.

\begin{figure*}
    \centering
    \includegraphics[width=1\linewidth]{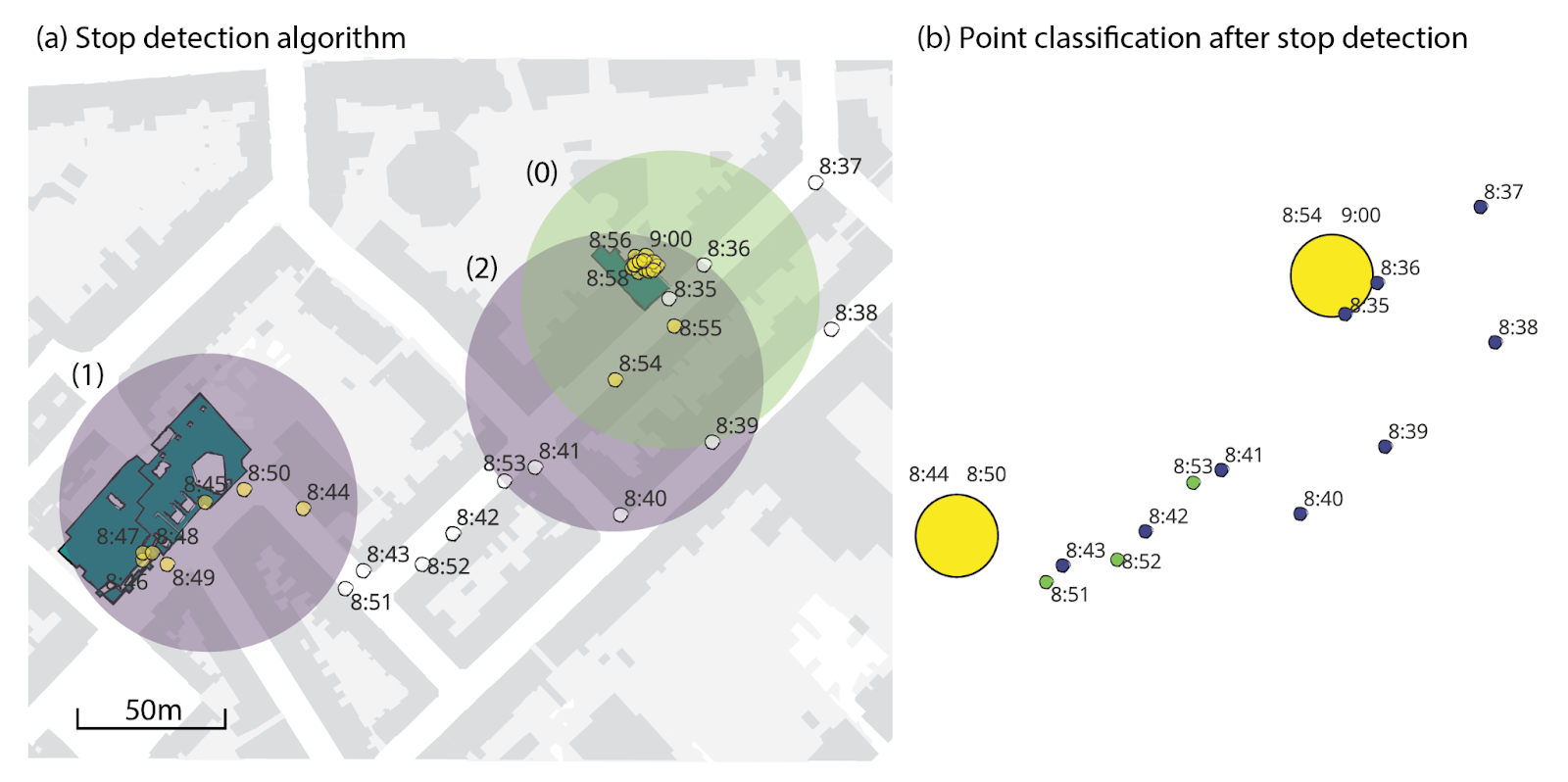}
    \caption{Derivation of the diary stop/trip format on the basis of the raw locations, using the algorithm from \cite{Ye2009-dv} in which time and radius parameters are used to cluster the timestamped geolocation data. In this method, a stop is defined as a length of time (here 5 minutes), during which all points fall within a given distance (here 50 meters) of the originating point. a) purple circles represent stops derived with this algorithm, corresponding generally to locations (1) “Son’s School” and (2) “Home” from the diary-formatted trips in Figure~\ref{fig:diaryvsraw} a). The green circle labeled (0) lacks a sufficient number of locations for the stop to be detected and is therefore missing as an originating stop for the first trip. b) points are classified according to the algorithm. Geolocations contributing to stops (1) and (2) are clustered for deriving further variables concerning these stops. Points not occurring during stops are clustered temporally to derive trip-related variables. Blue points represent the first trip between stops (0) and (1), and green points represent the return journey between stops (1) and (2).}
    \label{fig:stopstart}
\end{figure*}

In practice, the continuous geolocation stream of data in smart surveys are split into a series of trips (periods on the move) and stops (stationary periods). While the diary benefits from a respondent's own intuitive understanding of the trip/stop distinction, it is impossible to use intuition to define what is a 'trip': smart surveys need a decision rule to decide when a trip starts and ends. Figure~\ref{fig:stopstart} illustrates a spatiotemporal stop detection algorithm on the example data \citep{Montoliu2013-tb, Ye2009-dv}. After the stop algorithm is applied to the raw location data,the start and endpoints of trips and stops have to be precisely geocoded in a second step. This is a relatively straightforward procedure, that can be done by matching a latitude and longitude to an address or point of interest.  The two steps result in a processed dataset is produced as shown in Table \ref{tab:diarygeneration}. 
In a traditional diary, respondents provide the exact address of the start and end location of every trip directlt. They are more unlikely to provide an incorrect or approximate address than the reverse geocoding process, but more likely to leave this field blank. For this example, it is howveer possible to process data from mixed-mode and smary-survey data in such a way that data are quite comparable.

\begin{table}
    \small\sf\centering
    \caption{The state of the diary-generation process following each step. The number of rows, and trip start and end time are provided by stop detection. The address is provided by reverse geocoding.}
    \label{tab:diarygeneration}
    \begin{tabular}{p{1.5cm}p{1.75cm}p{.8cm}p{.8cm}p{1.25cm}p{1cm}}
    \toprule
        Where did you go? & Address & Trip start & Trip end & Transp. Method & Distance \\
    \midrule
        (1) & 203 Main St. & 08:35 & 08:43 & Walk & 210m\\
        (2) & 4 Church Ln. & 08:51 & 08:53 & Walk & 210m\\
    \bottomrule
    \end{tabular}
\end{table}

Apart from start and end-locations, travel surveys typically aim to also estimate  the  transportation method, label, and distance of every trip.
For transportation method, respondents in diary studies should report all methods (e.g., walking, car, train) they use per trip. In diaries, respondents accurately report the mode of transportation used for the largest leg of their journey, but often exclude transitory steps into and out of these primary modes, leading to underreporting \citep{Axhausen1995-dn}. Although identification of transportation modes from sensor data has become quite advanced, respondents still outperform algorithms in classifying mode of transportation. The increased granularity of the smart survey data would offer the capacity to segment trips into individual legs corresponding to differing transport modes, but the lack of a corresponding field in the non-smart survey means this additional information must be lost in combining the two modes into a single data set. Labeling stops provide a complementary example in which enriching information from a traditional survey may be lost during data integration. Raw data captured by smart surveys cannot be reliably enriched to provide a purpose or contextual label for stops. Respondent feedback is necessary for labeling contexts outside "home" or "work", but transport mode can to a large degree be inferred from sensor data alone \citep{Bantis2017-xs, zhao2015stop, sadeghian2021review}. In Table \ref{tab:diarygeneration}, an algorithm like that in \cite{Bantis2017-xs} has been applied to the example data to establish the primary mode of transportation during the two trips.

Finally, in both traditional diaries and smart surveys, algorithms are needed to calculate travel distance per trip. In a smart survey, travel distance can be reliably estimated on the basis of the exact route someone traveled. In a traditional diary survey, respondents are often asked to estimate how far the distance was that they traveled, or distance is inferred from the most logical route between the start and end-locations \citep{lamondia2016shifts,otpr}.
In smart surveys, trip distance can be similarly  calculated based on start and end-locations, but a much better way is to map the actual travel route and calculate distance traveled on the basis of that route.  For the example data, Open Trip Planner was used to calculate the walking distance between the stops given in Table \ref{tab:diarygeneration}. This process will lead to an underestimate of the actual distance of the first trip, in which a longer route was actually taken. 

If the goal is to harmonize data between the diary study and smart survey so that can be integrated in one dataset as in Table \ref{tab:diarygeneration2}, our example shows that compromises have to be made. Especially when it comes to travel distance and travel mode, diary surveys and smart surveys differ fundamentally in the way they do measurements. Harmonizing the data brings the benefit of analytical ease, but comes at the price of comprising measurement quality. 

\begin{table}
    \small\sf\centering
    \caption{The state of the diary-generation process following each step in our artifial example. The number of rows, and trip start and end time are provided by stop detection. The address is provided by reverse geocoding.}
    \label{tab:diarygeneration2}
    \begin{tabular}{p{.5cm}p{.8cm}p{1.75cm}p{1.75cm}p{.8cm}p{.8cm}p{1.25cm}p{1cm}}
    \toprule
        Day & Mode & Where did you go? & Address & Trip start & Trip end & Transp. method & Distance \\
    \midrule
        1 & Diary & Son's school & 203 Main St. & 08:35 & 08:50 & Walk & 210m \\
        1 & Diary & Home         & 4 Church Ln. & 08:50 & 09:00 & Walk & 210m \\ 
    \midrule
        2 & App   & -            & 203 Main St. & 08:35 & 08:43 & Walk & 210m \\ 
        2 & App   & -            & 4 Church Ln. & 08:51 & 08:53 & Walk & 210m \\ 
    \bottomrule
    \end{tabular}
\end{table}

\section{Multisource integration} \label{sec:multisource}

Conceptually, multi-source integration treats the different data collection modes as if they were entirely separate datasets which can then be analyzed using the methodology arising from multi-source statistics \citep{Waal2020-ii}. 

In multisource integration, each data source maintains its full structure because integration occurs at the end of the data analysis stage rather than beforehand. This model-first approach necessitates upfront consideration of the outcome statistics of interest. As an example, consider the calculation of the average distance traveled during a walking trip. 
In multi-source statistics, we first determine each source's unique contribution to total error reduction. First, how complete are data in either source? The smart survey may have less overall underreporting of trips, while the diary survey might offer more complete labeling of trip purpose. Individually, both modes may underestimate average trips—the diary due to day-level missing data and the smart survey due to trip-level omissions. 

Second, what is the quality of measurements? In the diary survey, we are limited to the start-end address formats of individual trips, and we have to calculate distance based on the most logical route traveled. In the smart survey, we calculate distance based on the exact route followed using GPS coordinates between the start- and end-location of a trip.

\begin{figure}
    \centering
    \includegraphics[width=0.5\linewidth]{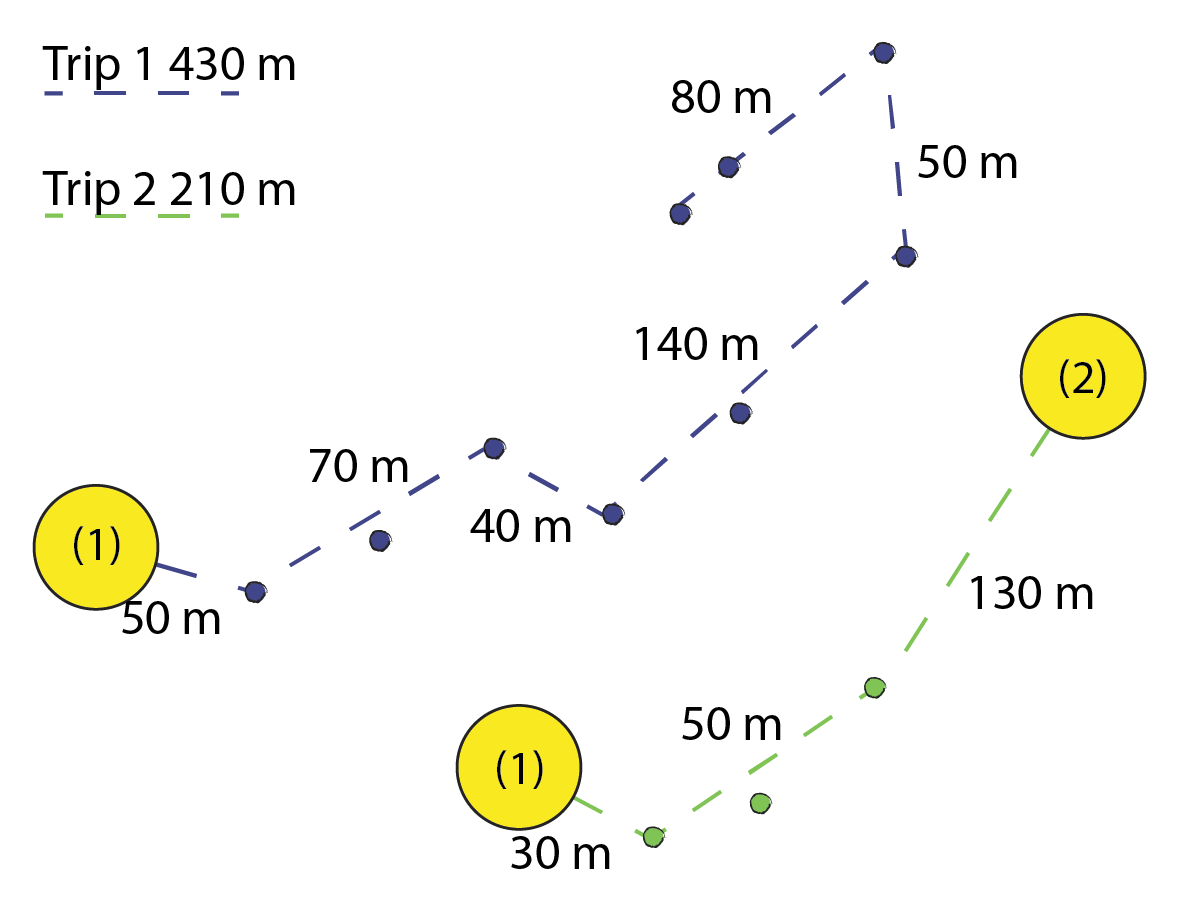}
    \caption{Calculating the trip distance for both example trips on the basis of the geolocations.}
    \label{fig:tripdist}
\end{figure}

Revisiting the two-trip example from Sections~\ref{sec:twotypes} and \ref{sec:mixedmode}, we calculate a distance of 430 meters for Trip 1 and 210 meters for Trip 2 on the basis of the recorded geolocations, as shown in Figure~\ref{fig:tripdist}. Comparing this to Table~\ref{tab:diarygeneration2} with an address-derived distance of 210 meters, we overestimate Trip 1 by 220 meters. In Section~\ref{sec:mixedmode}, we traded increased total measurement error for a decrease in relative mode measurement bias with respect to the diary.  However, we can potentially use the smart survey data to first calculate the ratio between the actual distance and inferred distance, and use that ratio to correct our traditional survey estimates \citep{Dalton2015-mh}. As with calibration in Section~\ref{sec:mixedmode}, this allows for estimation of mode measurement differences as well as uncertainty, but unlike with calibration, the relationship between the benchmark and non-benchmark mode is parameterized directly.

A more realistic example might be the prediction of travel mode preferences. Using the survey diary data, we can use the trip statistics such as distance and reported travel mode to build a travel-mode prediction model given variables such as trip length, and the nature of the start- and end locations. In a smart survey, the continuous geolocations provide a rich layer of spatiotemporal information that can be used to augment an integrated travel-mode prediction model that uses for example spatial covariates directly \citep{wu2022travel}, such as environmental features or speed at different points of the trip. Although the two different datasets are processed independently, they will still result in two databases that can to some degree be integrated. Depending on the mode that was used for collection, the integrated database will have missing data (because statistics such as actual distance would not be available in the survey), but also complete variables. For the complete variables, there will still be measurement ifferences that directly depend on the mode of data collection. These can be estimated and corrected for by use of calibration, or by using imputation of counterfactual data.

In short, the idea of multisource-integration is to use the strengths and weaknesses of different data sources when combining them into outcome statistic. This practice has been common in transportation research for decades. Travel Mode Choice modeling is used to understand and predict how individuals decide among various transportation modes given a set of circumstances. The most common models are built on multiple data sources, including census data, transit smart card data, mobile phone data, traffic count data, and travel diaries \citep{Graells-Garrido2023-dv, Zhang2019-kz}. The transition towards smart surveys and away from traditional travel diaries to smart surveys within these models involves the abstraction of the single data source into multiple independent sources that work in tandem in the model. 

Multi-source data integration is very flexible, which allows for a more bespoke integration process. This can be a benefit when the modes have large measurement differences or when no mode is a sensible benchmark because both are error-prone. On the other hand, it requires explicit modeling of a specific research question and doesn't offer the analysis flexibility of a single, generalized data set. It also complicates producing population level statistics, such as the the average number of trips, or distance traveled. 

\section{A decision framework}\label{sec:decision}

In comparison to traditional survey diaries, smart surveys can offer improvements to data quality. However, these improvements generally lead to differences in measurement and response patterns between smart- and traditional diary surveys. Finding ways to navigate these issues is critical to the success of smart surveys. In this section, we hope to provide researchers guidance on how to select between them, and to outline key survey design considerations that should be informed by the chosen approach.

In this paper, we have distinguished two distinct approaches to integration: a mixed-mode approach and multisource approach. Briefly, the mixed-mode strategy prioritizes outcome alignment, and its simplicity makes it preferable where measurement differences are expected to be small. The multisource strategy privileges inherent mode differences and is preferable when measurement differences between the two methods increase, although the flexibility increases its complexity. 

The mixed-mode approach has many benefits. Is easier to implement and allows for a continuation of analysis processes that have been used using data from traditional diary studies. Data can be plugged into existing models or compared longitudinally without any new model adaptation. Methodology exists for supporting mixed-mode analysis as long as it is possible to account for the various causes of structural difference between modes \citep{Buelens2018-eb, Clarke2022-mr}, and the approaches are particularly well-suited to situations in which one mode can be taken as a more accurate baseline against which the other may be calibrated \citep{Buelens2017-oa}.

The multisource approach offers flexibility in research design and allows researchers to exploit the benefits of each source without requiring alignment on a per-variable basis. This can be accomplished on a micro-level in which a simulated dataset is constructed on the basis of both sources, or at a macro-level in which the contribution of each source is generally explicitly parameterized \citep{Waal2020-ii}. Maintaining the structure of each source is well-suited to situations in which the new mode captures at least some data with a fundamentally different structure or level of granularity, as is often the case with smart surveys involving real-time or passive data. Delaying source integration until the modeling phase can also allow for otherwise difficult-to-incorporate measures to contribute to the model \citep{von-Behren2024-bz}.

\begin{figure}
    \centering
    \includegraphics[width=0.75\linewidth]{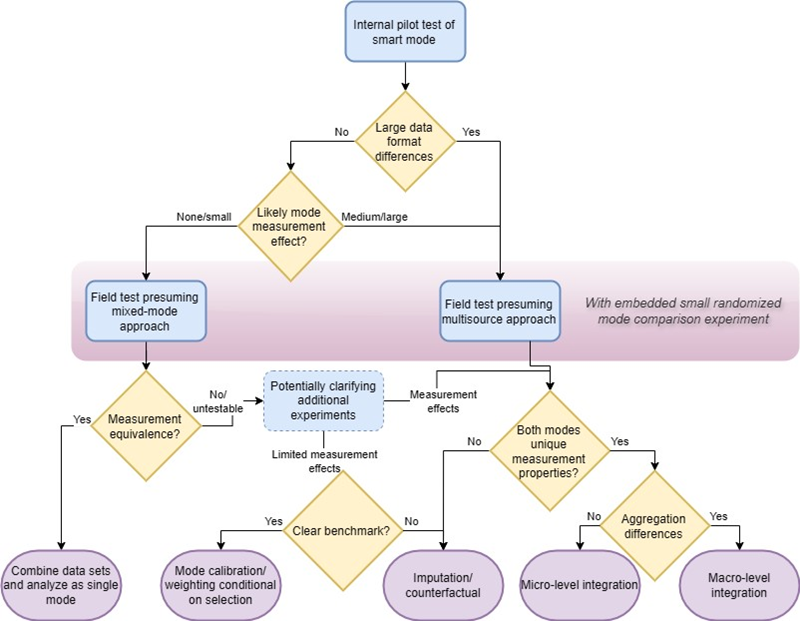}
    \caption{A decision framework for choosing between mixed-mode integration (left) and multisouce integration (right), both before (above), and after (under), a field test.}
    \label{fig:flowchart}
\end{figure}

Figure~\ref{fig:flowchart} provides a simplified flow chart that can help guide a decision on how to integrate data. First, rResearchers should be attentive to what the raw data in both a diary and smart survey measure and how this aligns with the desired measure (the construct validity). Perhaps one method is clearly superior in terms of validity to another, perhaps both methods have different measurement problems. What matters for the decision of how to integrate the two datasets is whether the measures are inherently different.

The best way to test this is by using small-scale pilot data. Collecting data from a small-scale pilot, often as part of the user experience (UX) tests, can be used to assess the bounds of an overall mode effect at an early stage. For example, a marked increase or decrease in total distance traveled within a travel study or a suspicious distribution of activity time in a time use survey would be evident at this point to a researcher. While these aspects are often flagged as issues with the application or the UX design, they can also serve as an early warning sign that the eventual data will be different enough to warrant designing with respect to mode effects.

Where limited mode measurement differences are expected, it can be helpful to embed small randomized mode comparison experiments into the field test to gauge the size of the differences better. Where substantial mode measurement effects are expected, doing experiments becomes critical to do so, as it will help provide the necessary information for disentangling measurement and representational differences \citep{Revilla2013-dt, De_Leeuw2018-pd, Vannieuwenhuyze2014-tu}. 

If measurement differences are assumed or tested to be small or negligible, the preferred approach is to use a mixed-mode like integration approach, where the data from one mode is processed in such a way that it resembes the data structure of the other dataset, and can be analysed jointly. Because smart survey data are likely to be much more granular, this often involves aggregating or processing the smart data to resemble the survey data.

If measurement differenes are expected, or found to be large, some multisource is needed. In the most extreme version of this, data are kept completely separate, and only the final outcome statistics (e.g. the mean distance traveled per person per day), are integrated (macro-integration). Data may also be integrated within one dataset with varying degrees of overlap between variables, depending on the number of variables in the study where measurement differences are small or large.

There are some potential challenges with adopting this decision framework at the outset of the survey design process. First, researchers often find themselves in a situation in which they expected smaller differences between modes than were encountered in practice, leading to initial design decisions made with mixed-mode analyses in mind that prove non-viable at the analysis stage. Here, some small experimental designs may be of use in establishing sufficient additional information to support the analysis. Re-interview designs in which a subset of participants are recontacted to participate under the alternative mode are one such option \citep{Klausch2017-vs, Schouten2024-vm}, and small-scale post-hoc experiments using e.g. cognitive interviewing \citep{Campanelli2015-xj} another.

Secondly, it may still be unclear whether a mixed-mode or multisource approach would be preferable, for example in a situation where the data are collected at the same level of aggregation or there is a desire to maintain the existing data format, but where the mode measurement effect is substantial enough to reduce the set of mixed-mode options for integration. Many analytical techniques also straddle this boundary, for example by making use of statistical matching or multiple imputation as a mechanism for aligning datasets, or incorporating existing methodology into a Bayesian design \citep{Boeschoten2018-ui, Boeschoten2016-mp, Kolenikov2014-wr, Yu2024-xs}.

Perhaps the largest challenge is the current lack of clear guidance on implementation. Although some disciplines like transportation and mobility have decades of work underpinning the theoretical properties of the multisource models, many do not. In this case, it may be appropriate to analyze both modes independently for a time and to defer integrated analyzes to a future moment. 

Ultimately, researchers who plan on integrating smart and non-smart surveys will need to make many choices along the way, often on the basis of less data than they would want. Nevertheless, an early decision to focus either on a mixed-mode or a multisource approach design should be made explicitly and allowed to inform all design decisions. 

\section{Discussion}\label{sec:discussion}

The decision to integrate smart and traditional surveys introduces inherent complexities in the design, raising the question of whether there is sufficient benefit to fielding traditional surveys anymore. Considering the rapid uptake of new forms of technology, this may be a viable alternative for surveys in which no historic comparison is necessary. The literature contains many such smart surveys deployed alone to good effect in populations unlikely to suffer from selection effects, and where historic precedent was not a concern, but these factors form a limiting concern for many researchers. More generally, this discussion can be seen as a restatement of the unified mode design/method maximization discussion in which, despite initial proponents on both sides, the uni-mode paradigm was eventually considered best practice \citep{de-Leeuw2018-hk}. Indeed, when a mixed-mode approach is feasible, we recommend that the unified design principles proposed by \cite{Dillman2014-yd} be followed. Although similar in its theoretical basis, the resurfacing of this question in the context of smart surveys is unsurprising, as smart features are designed to mode-inherent differences that cannot be overcome by questionnaire design.

Combining smart and traditional surveys involves some inherent challenges. Randomizing between modes, at least for some portion of the sample, is recommended in order to provide a framework for establishing the portion of the mode effect due to selection, but this may not always be possible. Regardless of whether a mixed-mode or multisource approach is taken, valid inference depends on accurate estimation of differences in coverage and response, and non-random mode assignment increases these differences. While propensity score adjustment on the basis of demographic variables may explain some selection bias, it is unlikely to be sufficient \citep{Kibuchi2024-tn}.

Both the multisource and mixed-mode approaches have inherent limitations only partially addressed by the alternative. Multisource analysis is more complex, and complex models that are not completely understood or properly specified can be misleading, doing more harm than good \citep{Chesher2011-vo}. This is made more difficult by the current lack of research providing theoretical underpinnings for many multisource estimation procedures. Two recent ESSnet projects have been undertaken with the goal of providing a thorough overview of multisource and mixed-mode statistics. These are respectively \cite{Unknown2019-gy} and \cite{Signore2019-jk}. Though neither project explicitly focuses on smart survey integration concerns, the deliverables contain guidelines and recommendations that are highly applicable for researchers designing along either approach. 

Of course, the two broad approaches presented in this paper do not cover all possible situations and arrangements of smart and non-smart surveys. Instead, we hope that they provide researchers with a starting point, and a set of analytic considerations that will guide a larger discussion on when to use which approach. As smart surveys continue to develop, neither of these two approaches may be sufficient. As a general rule, the smarter the survey becomes, the larger the differences in measurement become. This may mean that integration of smart and non-smart surveys becomes too complex to manage through mixed-mode methods. Alternatively, evolving frameworks for passive and sensor data may remove the need for many researchers to develop complex models on the basis of the raw data by providing for statistic-specific queries. 

Neither mixed-mode nor multisource systems are new, but smart surveys, at least at the moment, straddle an unclear boundary between the two, requiring us to update our knowledge on the bounds and extents to which these modes can impact data quality. There is a real need for comparative studies reporting on studies that have fielded both in experimental settings. 

Often, opting for the more conservative mixed-mode approach may be the correct call for a national statistical institute fielding a smart survey for the first time, with a goal of transitioning to a forward-looking multisource approach as the field continues to evolve. This means that retaining and analyzing the new data forms in their raw format is essential, even if the decision is made to generate statistics on the basis of aggregation to flat files. This approach allows for future methodological advances to be assessed longitudinally, and begins to pave the way for future methodological advances. By embedding flexibility into data collection strategies, designs can be futureproofed while maximizing the potential of both traditional and smart surveys. 

% Add something about how it's applicable to time use (e.g. the ten minute slot thing)
% We know people brush teeth once or twice a day but the diary data doesn't reflect this because of the ten minute blocks, so people either round up or don't report it or it aggregates into some larger category. What do you do with this if you want to integrate it? Either round up
% Impute the data reports on the time use, then we have the division of the estimates more granular, then we can get to an inferred time use per person and we can infer that people brush their teeth
% Miss a lot of very short activities, could theoretically have in the smart, various ways of integrating, won't go into it

%12Similarly, when smart surveys are introduced as a mode in longitudinal surveys, changes over time may be confounded with mode effects, and may require specific experiments within subgroups to distinguish group differences, estimates of change, and mode effects \citep{Jackle2017-qa, Cernat2021-cx}. 

%14The approach chosen depends on the research question, but researchers are often limited by the way things have been done previously. This may have a tendency to bias data collection efforts towards conservatism, limiting innovation in the field. On the other hand, collecting data assuming that the questions will present themselves after the fact is expensive and violates an unspoken agreement between researcher and the respondent that their efforts are necessary.

\bibliographystyle{SageH}
\bibliography{dipaper}

%\section{Support for \textsf{\journalclass}}
%We offer on-line support to participating authors. Please contact
%us via e-mail at \dots
%
%We would welcome any feedback, positive or otherwise, on your
%experiences of using \textsf{\journalclass}.

%\section{Copyright statement}
%Please  be  aware that the use of  this \LaTeXe\ class file is
%governed by the following conditions.

%\subsection{Copyright}
%Copyright \copyright\ \volumeyear\ SAGE Publications Ltd,
%1 Oliver's Yard, 55 City Road, London, EC1Y~1SP, UK. All
%rights reserved.

%\subsection{Rules of use}
%This class file is made available for use by authors who wish to
%prepare an article for publication in a \textit{SAGE Publications} journal.
%The user may not exploit any
%part of the class file commercially.

%This class file is provided on an \textit{as is}  basis, without
%warranties of any kind, either express or implied, including but
%not limited to warranties of title, or implied  warranties of
%merchantablility or fitness for a particular purpose. There will
%be no duty on the author[s] of the software or SAGE Publications Ltd
%to correct any errors or defects in the software. Any
%statutory  rights you may have remain unaffected by your
%acceptance of these rules of use.

\end{document}